\documentclass[referee,sn-nature]{sn-jnl}
\usepackage{graphicx}%
\usepackage{multirow}%
\usepackage{amsmath,amssymb,amsfonts}%
\usepackage{amsthm}%
\usepackage{mathrsfs}%
\usepackage[title]{appendix}%
\usepackage{xcolor}%
\usepackage{textcomp}%
\usepackage{manyfoot}%
\usepackage{booktabs}%
\usepackage{algorithm}%
\usepackage{algorithmicx}%
\usepackage{algpseudocode}%
\usepackage{listings}%
\usepackage[english]{babel}
\usepackage[autostyle]{csquotes}
\usepackage[normalem]{ulem}
\usepackage{hyperref}
\usepackage{caption}
\usepackage{subcaption}
\usepackage{amsfonts}
\usepackage[T1]{fontenc}
\usepackage{pdflscape}
\usepackage{upgreek}

\RequirePackage{fix-cm}

\begin{document}

\title[Article Title]{High-speed synchrotron X-ray imaging of melt pool dynamics during ultrasonic melt processing of Al6061}

\author[1]{\fnm{Lovejoy} \sur{Mutswatiwa}}

\author[1]{\fnm{Lauren} \sur{Katch}}

\author[1]{\fnm{Nathan John} \sur{Kizer}}

\author[1]{\fnm{Judith Anne} \sur{Todd}}

\author[2]{\fnm{Tao} \sur{Sun}}

\author[3]{\fnm{Samuel James} \sur{Clark}}

\author[3]{\fnm{Kamel} \sur{Fezzaa}}

\author[4]{\fnm{Jordan} \sur{Lum}}

\author[4]{\fnm{David Matthew} \sur{Stobbe}}

\author[5]{\fnm{Griffin} \sur{Jones}}

\author[5]{\fnm{Kenneth Charles} \sur{Meinert Jr.}}

\author[1]{\fnm{Andrea Paola} \sur{Argüelles}}

\author*[1]{\fnm{Christopher Micheal} \sur{Kube}}\email{kube@psu.edu}

\affil[1]{\orgdiv{Engineering Science and Mechanics}, \orgname{The Pennsylvania State University}, \orgaddress{\street{212 Earth and Engineering Sciences Building}, \city{University Park}, \postcode{16802}, \state{PA}, \country{USA}}}

\affil[2]{\orgdiv{Materials Science and Engineering}, \orgname{University of Virginia}, \orgaddress{\street{Wilsdorf Hall, 395 McCormick Road}, \city{Charlottesville}, \postcode{22904}, \state{Virginia}, \country{USA}}}

\affil[3]{\orgname{X-ray Science Division}, \orgaddress{\street{Advanced Photon Source, Argonne National Laboratory}, \city{Lemont}, \state{IL}, \postcode{60439}, \country{USA}}}

\affil[4]{\orgname{Lawrence Livermore National Laboratory}, \orgaddress{\street{7000 East Ave}, \city{Livermore}, \postcode{94550}, \state{CA}, \country{USA}}}

\affil[5]{\orgdiv{Applied Research Laboratory}, \orgname{The Pennsylvania State University}, \orgaddress{\street{230 Innovation Blvd}, \city{University Park}, \postcode{16802}, \state{PA}, \country{USA}}}

\abstract{Ultrasonic processing of solidifying metals in additive manufacturing can provide grain refinement and advantageous mechanical properties. However, the specific physical mechanisms of microstructural refinement relevant to laser-based additive manufacturing have not been directly observed because of sub-millimeter length scales and rapid solidification rates associated with melt pools. Here, high-speed synchrotron X-ray imaging is used to observe the effect of ultrasonic vibration directly on melt pool dynamics and solidification of Al6061 alloy. The high temporal and spatial resolution enabled direct observation of cavitation effects driven by a 20.2 kHz ultrasonic source. We utilized multiphysics simulations to validate the postulated connection between ultrasonic treatment and solidification. The X-ray results show a decrease in melt pool and keyhole depth fluctuations during melting and promotion of pore migration toward the melt pool surface with applied sonication. Additionally, the simulation results reveal increased localized melt pool flow velocity, cooling rates, and thermal gradients with applied sonication. This work shows how ultrasonic treatment can impact melt pools and its potential for improving part quality.}

\maketitle

\section*{Introduction}
Laser-based metal additive manufacturing (AM), a three-dimensional printing technique, can manufacture single components and structures with highly complex geometries, functionally graded alloys \cite{Zhang2019}, tailored microstructures \cite{Dehoff15}, and enhanced mechanical properties \cite{Lewandowski2016}. However, for most alloys, thermal cracking, porosity, and columnar grains \cite{Arısoy2019} reduce mechanical properties and prevent the widespread adoption of AM parts \cite{Sames2016}. Establishing techniques for influencing solidification toward grain refinement could lead to parts with better mechanical properties and, ultimately, improve the reliability and quality of AM components \cite{Mohammadpour2020}. The variation of AM process parameters, such as laser power, scan speed, and energy density \cite{Okugawa22} allows control of thermal gradients and cooling rates, resulting in location-specific microstructural refinement \cite{Dehoff15}. However, process parameter optimization can be challenging, especially for alloys that are difficult to print. In addition to process parameter adjustment, inoculants can be added to the AM process to promote heterogeneous nucleation in the melt pool, resulting in grain refinement \cite{Martin2017}. However, inoculants unavoidably change the chemical composition of the material, which can impact the mechanical strength of AM components \cite{Spierings2016}. In addition, inoculants can cause inclusions due to settlement and agglomeration \cite{Xu2021}.

\medskip

\noindent Other techniques for solidification control can be achieved by applying external fields such as electromagnetic \cite{Wang21}, mechanical \cite{Colegrove2017}, or acoustic \cite{ESKIN1994} fields. In casting, electromagnetic fields were reported to increase cooling rates \cite{Wang08}, which resulted in reduced alloying element segregation and a more homogeneous macrostructure \cite{WangB18}. Low-frequency mold vibration also succeeded in solidification manipulation during casting, resulting in a refined as-cast grain structure \cite{Kisasoz17}. The application of high-intensity ultrasound on solidifying metals for molten metal processing during welding resulted in grain refinement and improved weld joint strength \cite{Krajewski12}. Nonetheless, using these techniques in laser-based metal AM is challenging because of the short length and time scales involved in melt pool dynamics and solidification \cite{Yang2021}. 

\medskip

\noindent Following the work of Eskin \cite{Eskin01} and Abramov \cite{Abramov86}, and applying successful grain refinement techniques in welding \cite{Sui22}, Todaro et al. \cite{Todaro20} recently demonstrated that high-intensity ultrasound can promote columnar to equiaxed grain transitions (CET) in laser AM fabricated Ti-6Al-4V and Inconel 625. As a result, components built with a fine, equiaxed grain structure exhibited increased yield and tensile strengths. One form of ultrasonic melt processing in AM involves laser metal powder consolidation on a substrate vibrating at ultrasonic frequencies (i.e. sonicated substrates). An applied ultrasonic frequency of $20$ kHz on an AM-fabricated 316L stainless steel plate resulted in a noticeable decrease in grain sizes and an increase in random grain orientations \cite{Todaro2021}. Similarly, a reduction in mechanical property anisotropy and grain refinement along the build direction in wire arc AM was recently observed after ultrasonic treatment \cite{Feilong2023}. Ivanov et al. \cite{Ivanov21} and Yoon et al. \cite{Yoon2022} leveraged high-frequency pulsed laser irradiation to introduce high-intensity ultrasonic waves in the melt pool, resulting in microstructural refinement. Wang et al. \cite{Wang2020} used ultrasonic vibration-assisted AM to fabricate Inconel 718 parts and investigated the influence of four ultrasonic frequencies (i.e. $0$, $25$, $33$, and $41$ kHz) on microstructural refinement and mechanical properties. While ultrasonic melt processing at $25$ kHz increased mechanical strength, the use of higher ultrasonic frequencies was observed to increase porosity and hardness. Wang et al. \cite{Wang2020} elucidated the effects of frequency, yet the effect of other ultrasonic wave parameters, such as vibration amplitude and acoustic intensity, on grain refinement, remained unclear.

\medskip

\noindent The observed microstructural refinement in AM ultrasonic melt processing reported in the literature is hypothesized to result from increased nucleation rates and sites caused by acoustic cavitation and streaming induced in the melt pool. Acoustic cavitation and streaming have been suggested to compete with Marangoni convection, recoil pressure, and surface tension forces in the melt pool \cite{Leung2018}, influencing solidification rates and thermal gradients and promoting columnar to fine equiaxed grain transitions \cite{Eskin19}. Cavitation was observed in high-speed synchrotron X-ray imaging experiments within a controlled casting with ultrasonic treatment by Wang et al. \cite{WangB18}. They observed acoustic cavitation bubbles imploding in a Bi-$8\%$Zn alloy on the solid-liquid interface, causing fragmentation of the solid phase in the mushy zone. Moreover, acoustic streaming was observed to disperse solid particles in the liquid, which have been reported to later act as solidification nuclei \cite{Wang17}. In AM, however, the melting and solidification processes occur rapidly, presenting spatial and temporal resolution challenges in direct cavitation observation. In their study focused on observing grain refinement mechanisms in ultrasound-assisted AM, Ji et al. \cite{Ji2023} stated that because of extremely high temperature, opacity, and short survival time, it is hardly possible to directly observe the process of ultrasound effect on the molten metal pool in AM through experiments. While direct observation of dendrite fracture would be challenging, recent high-speed X-ray imaging of keyhole dynamics in AM \cite{Zhao2020} allows observation of cavitation bubbles directly during ultrasound-assisted AM.

\medskip

\noindent In this work, high-speed synchrotron X-ray imaging at the Advanced Photon Source, Argonne National Laboratory was used to capture acoustic cavitation in high-temperature, viscous, and opaque sub-millimeter scale melt pools within an Al6061 sample. Ultrasonic treatment was observed to alter keyhole morphology, which could potentially reduce or eliminate porosity generated from keyhole tip collapse, in addition to reducing dynamic keyhole instabilities. Ultrasonic treatment influenced bubble dynamics, causing pore migration toward the melt pool surface. The reported results demonstrated the existence and influence of cavitation on laser-generated melt pool dynamics during ultrasonic melt processing, which was previously hypothesized by Todaro et al. \cite{Todaro2021, Todaro20}, Feilong et al. \cite{Feilong2023}, and Wang et al. \cite{Wang2020}. The multiphysics Computational fluid dynamics (CFD) simulations using the Flow-3D\textcircled{\scalebox{0.7}{R}} platform showed an increase in melt pool flow velocity, thermal gradients, and cooling rates with applied ultrasonic treatment. This study provides direct evidence that acoustic cavitation effects are present in laser-generated melt pools and can be studied using high-speed X-ray imaging and CFD simulations. Thus, controlling acoustic cavitation, microstructure, and, henceforth, mechanical properties and part quality is now a closer reality \cite{Zhang2023}.

\section*{Results}

\begin{figure}[t]
    \centering
    \includegraphics[width=0.75\textwidth]{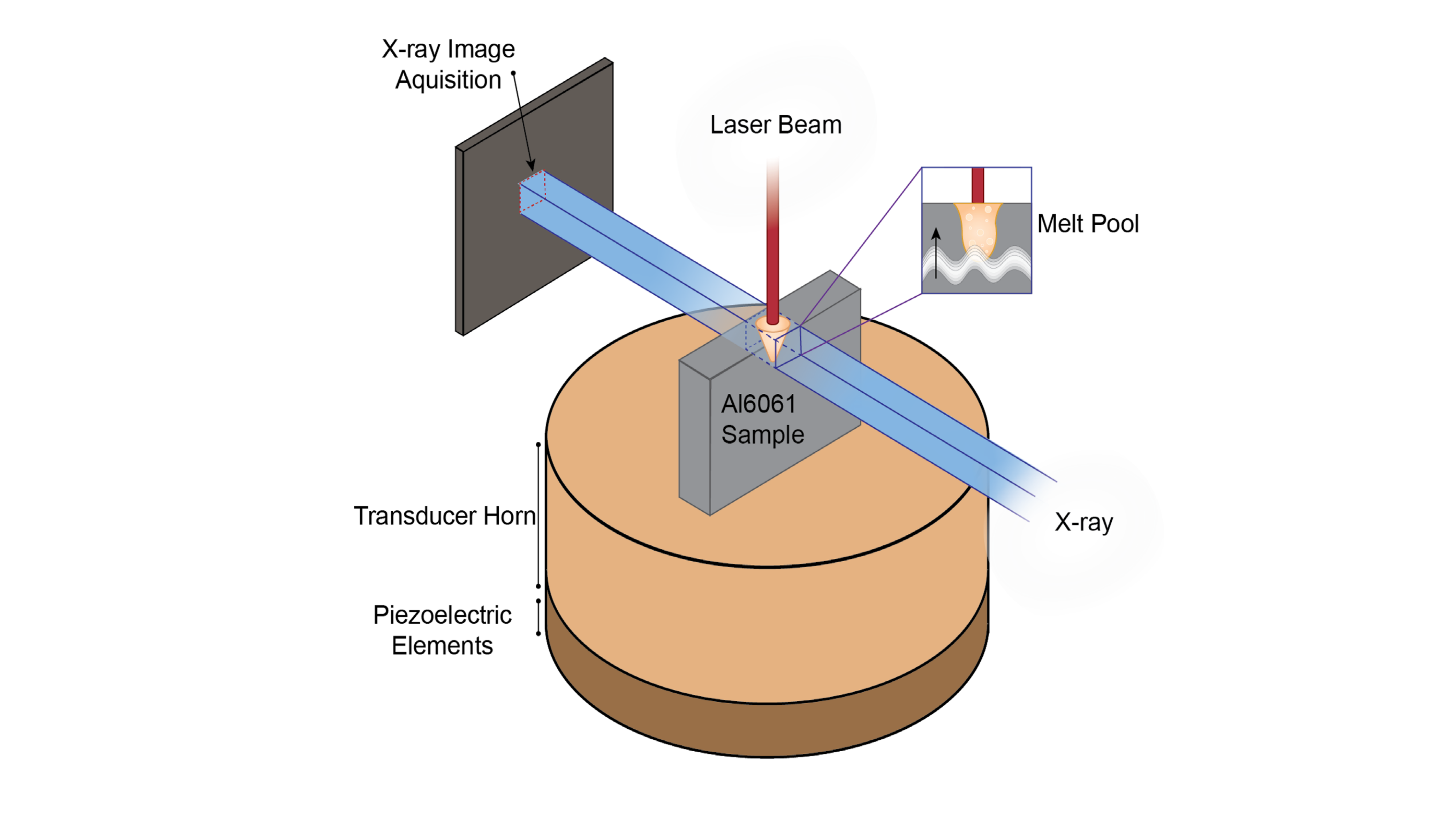}
    \caption{\textbf{Experimental setup.} Schematic diagram illustrating the experimental setup for high-speed X-ray imaging of melt pools on a vibrating substrate.}
    \label{fig: Experimental Setup Schematic diagram}
\end{figure}

\textbf{In-situ synchrotron X-ray imaging of acoustic cavitation in melt pools.} Figure \ref{fig: Experimental Setup Schematic diagram} shows the primary features of the experimental setup. 
The experiment consisted of a continuous-wave ytterbium fiber laser with user set powers ranging from $100$ to $560$ W, the high-speed X-ray imaging system (see details in \cite{WangB18}), and an Al6061 sample mounted vertically on top of a Langevin transducer driven at its lowest order extensional resonance frequency of 20.2 kHz. Single-pulse X-ray images were collected at a rate of $50$ kHz to observe melt pool dynamics, cavitation bubble dynamics, and solidification. X-ray computed tomography (CT) and electron backscattered diffraction (EBSD) were used to further characterize the pore structure ex-situ.

\medskip

\begin{figure}[t]
    \centering
    \includegraphics[width=0.75\textwidth]{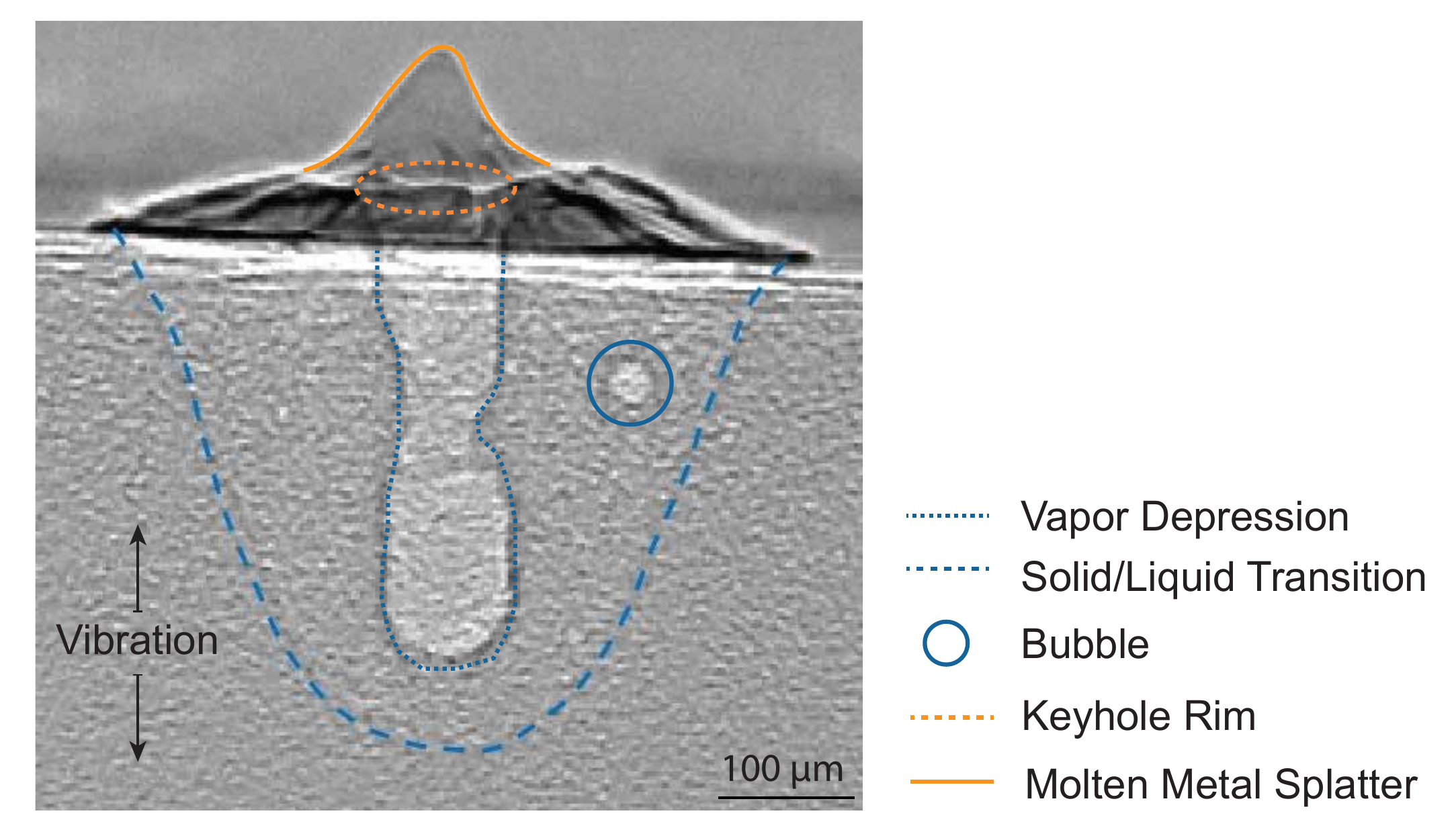}
    \caption{\textbf{Melt pool X-ray frame.} An X-ray frame showing the melt pool boundary at the solid/liquid interface, the keyhole or vapor depression morphology, the keyhole rim, hot spatter, a microbubble, and vibration direction. The video from which this frame was extracted is found in Supplementary Movie 1.}
    \label{fig: Representative single pulse X-ray}
\end{figure}

\noindent A representative X-ray image showing annotated melt pool features and vibration direction is shown in Figure \ref{fig: Representative single pulse X-ray}.
X-ray absorption and phase contrast allowed easy identification of the solid/liquid transition region, vapor depression area, and microscale bubbles from cavitation. Supplementary Movie 1 shows the entire single-point melt pool and solidification process when the 350 W laser is applied for $3.34$ ms without sonication. In addition, the video shows highly dynamic features such as bubble motion, melt pool size fluctuation, and keyhole initiation, growth, and fluctuation. The high spatial (i.e. $2$ $\upmu$m/pixel) and temporal (i.e. 50,000 frames per second) resolutions afforded by the high-energy synchrotron facility enabled direct quantifiable observation of the microscale bubble dynamics within the melt pool. The effect of the vibration could then be easily observed by conducting measurements with and without the active ultrasonic transducer. While the vibration was active, the X-ray imaging allowed direct measurement of the vibration amplitude of approximately $8$ $\upmu$m (more details on image processing and measurements are provided in the \enquote{Methods} section).

\medskip

\begin{figure}[t]
    \centering
    \includegraphics[width=1\textwidth]{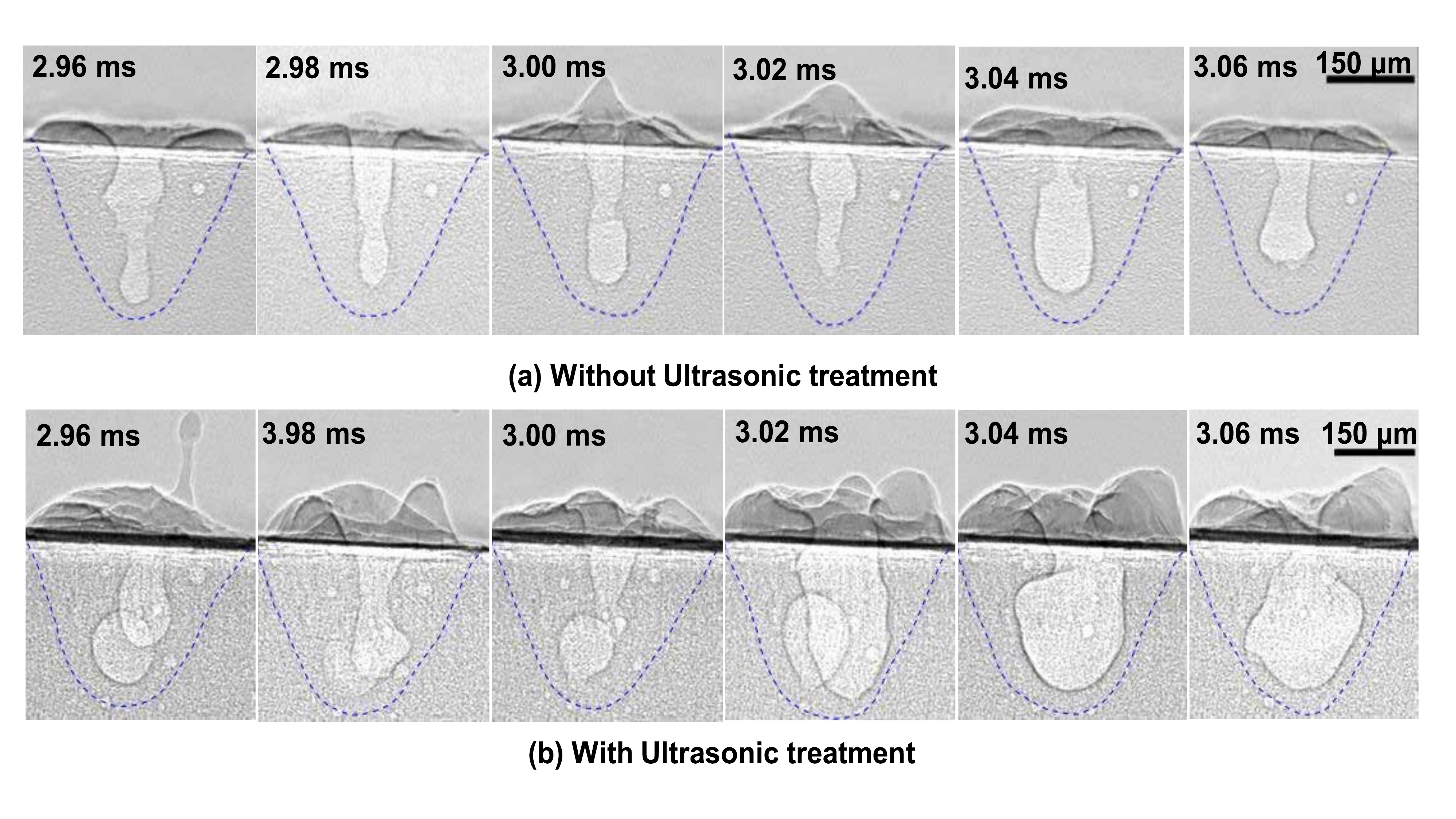}
     \caption{\textbf{ X-ray image sequences showing laser-generated molten Al6061 pools.} Melt pools (a) without and (b) with sonication. The six X-ray frames were taken at $0.02$ ms intervals, beginning at $2.96$ ms after the laser was turned on. The video from which these frames were extracted is found in Supplementary Movie 2.}
    \label{fig: X-ray frames 2.96 ms after laser turns on}
\end{figure}

\noindent Figures \ref{fig: X-ray frames 2.96 ms after laser turns on}(a) and
\ref{fig: X-ray frames 2.96 ms after laser turns on}(b) depict real-time X-ray image sequences of stationary laser-generated molten Al6061 pool dynamics without and with sonication, respectively. Supplementary Movie 2 is the associated high-speed videos containing the frames seen in Figures \ref{fig: X-ray frames 2.96 ms after laser turns on}(a) and (b). In Figure \ref{fig: X-ray frames 2.96 ms after laser turns on}(a) a narrow and deep vapor depression or keyhole can be observed in melt pools without sonication. Keyhole melt pools with these characteristics are known to be susceptible to keyhole porosity in AM when the tip of the vapor depression pinches off and forms a bubble \cite{Cunningham2019, Zhao2020}. Without sonication, bubbles were observed to settle at the bottom of the melt pool, where the solidification front could quickly freeze them, resulting in porosity. Figure \ref{fig: X-ray frames 2.96 ms after laser turns on}(a) also shows strong fluctuations in keyhole depth, which is a characteristic of keyhole instability \cite{Gan2021}. 

\medskip

\noindent The bubble density is shown to increase due to sonication as depicted in Figure \ref{fig: X-ray frames 2.96 ms after laser turns on}(b) and Supplementary Movie 2, proving the sonication leads to bubble nucleation in the liquid phase separate from the keyhole region. The bubbles in the melt pool with sonication rapidly nucleate, grow, oscillate, and sometimes implode, demonstrating cavitation bubble behavior. In addition, acoustic streaming effects were observed, where the molten metal flows in the vibration direction \cite{Balasubramani19, Yang22}. Sonication increased the average bubble diameter and promoted bubble migration towards the melt pool surface (Supplementary Movie 2). Bubbles with larger diameters were observed to implode at the melt pool surface, demonstrating degassing characteristics. In conventional AM, the melt flow-induced drag force dominates bubble dynamics \cite{Hojjatzadeh2019}. Based on the observed bubble dynamics in melt pools with sonication, it can be pointed out that bubble growth due to cavitation increases the buoyancy force, overcoming the drag force that usually traps pores \cite{Hojjatzadeh2019}, steering the bubbles toward the melt pool surface, and promoting degassing \cite{Eskin2018chp5}. In addition, we speculate that primary and secondary Bjerknes acoustic radiation forces may exist in the melt pool, facilitating bubble translation toward the melt pool surface and causing degassing \cite{Supponen2020}. The concentration of porosity toward the melt pool surface induced by sonication might be convenient in metal AM because the remelting between successive layers could eliminate the residual porosity from previous layers.

\medskip

\noindent Figure \ref{fig: X-ray frames 2.96 ms after laser turns on}(b) also shows a reduction in the keyhole depth fluctuations and an increase in the keyhole tip radius with sonication. These phenomena resulted in the elimination of the keyhole tip pinch-off porosity \cite{Zhao2020}. However, sonication was observed to eject molten metal from the melt pool, as shown in the X-ray frame at $2.96$ ms with sonication in Figure \ref{fig: X-ray frames 2.96 ms after laser turns on}(b). Further investigation on the influence of substrate vibration directions (i.e. in-plane or out-of-plane vibration) and vibration amplitudes and frequencies could help minimize potential spatter in laser-based AM with ultrasonic melt processing and will be explored in our future research.

\medskip

\begin{figure}[t]
    \centering
    \includegraphics[width=1.14\textwidth]{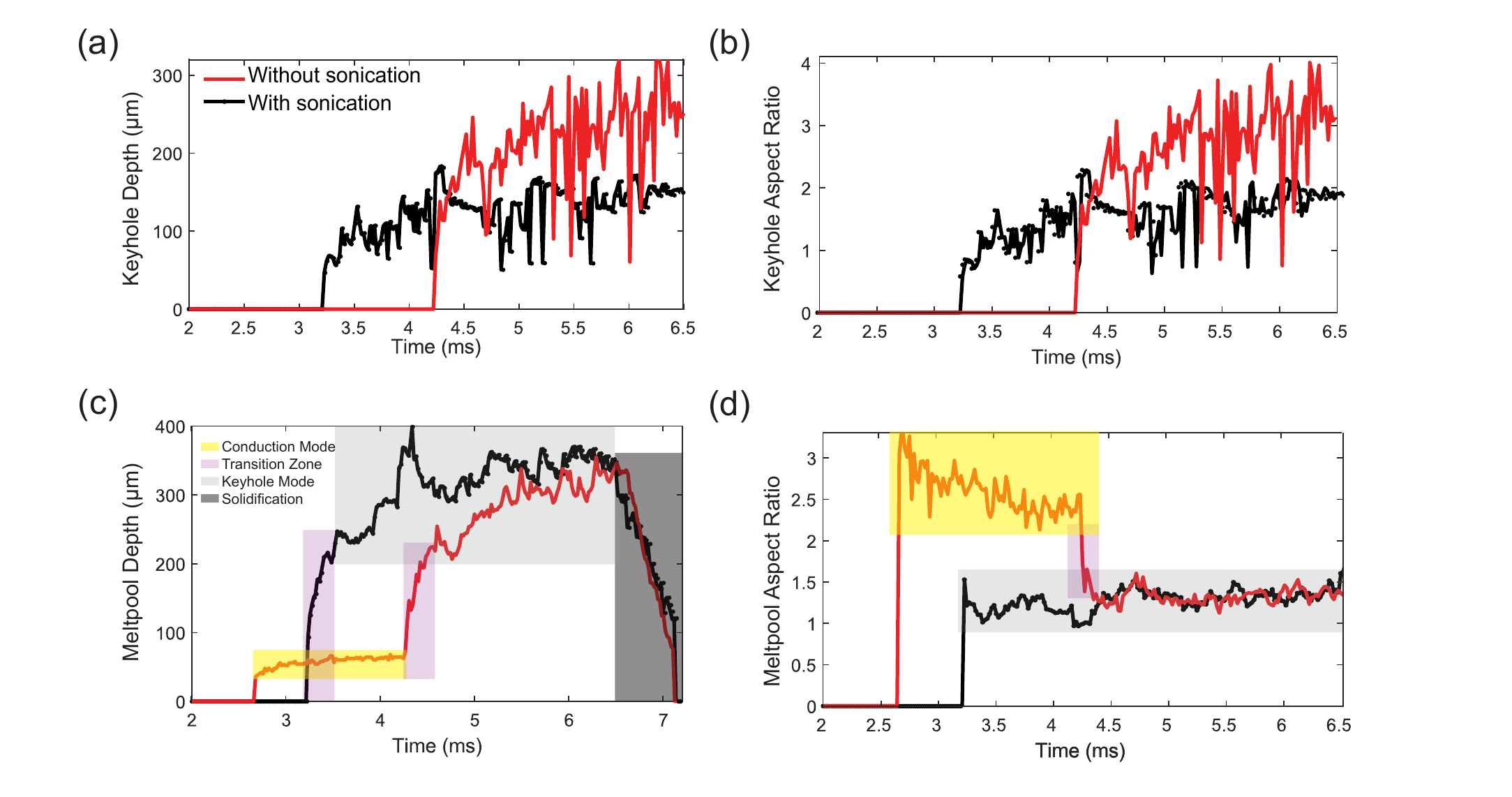}
    \caption{\textbf{Influence of ultrasonic treatment on melt pool and keyhole geometry.} (a) Keyhole depth, (b) Keyhole aspect ratios (keyhole depth divided by fixed laser beam diameter of $80$ $\upmu$m), (c) Melt pool depths, and (d) Melt pool aspect ratios (melt pool width divided by depth based on measurements from X-ray images) with and without sonication. Red plain line shows measurements without sonication while black line with circular markers shows measurements with sonication.}
    \label{fig: Melt pool geometry}
\end{figure}

\noindent \textbf{Influence of ultrasonic treatment on melt pool geometry and dynamics} The variations in the keyhole and melt pool depths, with and without sonication, are illustrated in Figure \ref{fig: Melt pool geometry}. The melt pool depths, keyhole depths, and melt pool widths  were measured from the point where sizable contrast difference between the liquid/solid and gas/liquid phases could be observed in the X-ray images. From Figure \ref{fig: Melt pool geometry} (a), it can be observed that the keyhole depth without sonication was larger
than the sonicated keyhole. Melt pool and keyhole depths were shown to fluctuate at constant laser power \cite{Guo2019} indicative of instabilities  \cite{Cunningham2019}. The depth fluctuations were quantified as one standard deviation about the mean of the measured depths. With sonication, the melt pool depth standard deviation was $66.3$ $\upmu$m, whereas it was $111.6$ $\upmu$m without. Similarly, the keyhole depth standard deviation was $31.6$ $\upmu$m compared to $57.6$ $\upmu$m with and without sonication, respectively. This indicates ultrasonic treatment reduces fluctuations, leading to more stable dynamics. Without sonication, the melt pool began in conduction mode as shown in Figure \ref{fig: Melt pool geometry} (c) from $2.8$ to $4.25$ ms, after which the melt pool transitioned into the keyhole mode. Conversely, with sonication, the melt pool started directly in keyhole mode. In both cases, the transition from conduction to keyhole mode occurred rapidly until stabilizing after about 5.5 ms.

\medskip

\noindent Keyhole morphology also plays a role in melt pool dynamics and defect formation during laser-based metal AM processes. Figure \ref{fig: Melt pool geometry} (b) shows the Keyhole aspect ratios calculated from measured depths divided by the 80 $\upmu m$ laser diameter \cite{Gan2021}. These results show higher aspect ratios in melt pools being sonicated. A high aspect ratio represents a deep and narrow keyhole with a needle-like tip, while a low keyhole aspect ratio represents a wide keyhole with an an observable tip radius. A deep and narrow keyhole traps laser beam reflections at the bottom, leading to a J-shaped keyhole in moving laser scenarios \cite{Zhao2020}, which are susceptible to keyhole tip pinch-off porosity \cite{Huang2022, Hojjatzadeh2019}. Therefore, ultrasonic treatment in metal AM can potentially eliminate one of the major keyhole porosity driving mechanisms by decreasing the keyhole aspect ratio and keyhole-tip radius. Figure \ref{fig: Melt pool geometry} (d) depicts the melt pool aspect ratio with and without sonication. In the absence of sonication, a high melt pool aspect ratio was observed when the melt pool was in conduction mode (i.e. from $2.7$ to $4.4$ ms) compared to the keyhole mode. There was not a significant difference in the melt pool aspect ratio due to sonication.

\medskip

\noindent Laser energy absorptivity is known to be influenced by melt pool and keyhole depths \cite{Khairallah2016}. Thus, the difference in melt pool geometries in ultrasonically treated melt pools relative to non-ultrasonically treated melt pools could result from the variation in the position of the laser focal point relative to the melt surface caused by the back-and-forth motion of the vibrating sample, promoting multiple laser beam reflections, resulting in improved laser energy absorptivity. This is possible at high vibration amplitudes to laser spot size ratios. However, in our case, an $16$ $\upmu$m peak-to-peak vibration amplitude and a laser spot size of $80$ $\upmu$m will not significantly influence laser energy absorptivity. Therefore, we speculate that the increased absorptivity could be due to the raised melt pool surface above the sample due to ultrasonic vibration causing the keyhole rim to rise while the recoil pressure keeps the bottom of the keyhole stationary. Hence, it results in deeper keyholes that promote multiple laser beam reflections on the vapor/liquid interface and increased absorptivity. In addition, the melt pool temperature could have increased because of bubble implosions, resulting in a larger melt region with applied ultrasound. Improved laser energy absorptivity and large melt pools are advantageous in metal AM to potentially reduce component build time. To investigate these claims further, CFD simulations were conducted to explain the impact of sonication on thermal gradients and cooling rates.

\medskip

\noindent \textbf{Multiphysics modeling of melt pool dynamics and solidification in ultrasound-assisted AM} High-speed X-ray imaging was able to provide real-time evidence of acoustic cavitation and melt pool dynamics in laser-generated melt pools driven by an external ultrasonic field. Additional insight into pressure distributions, thermal gradients, and cooling rate information is available through bridging the experiments with CFD simulations. In particular, CFD offers the ability to connect thermal properties to microstructural development. To further investigate the influence of ultrasonic treatment on solidification, we conducted multiphysics simulations of single-spot laser-generated melt pools with and without ultrasonic vibration using Flow-3D\textcircled{\scalebox{0.7}{R}}. Identical laser and ultrasonic parameters and substrate material used in the X-ray imaging experiments were adopted in the simulations. To reduce the simulation time, the laser duration was set to $0.8$ ms compared to $3.4$ ms in the experiments. The X-ray images were used to validate the simulations by directly observing melt pool and keyhole morphologies, cavitation bubbles, and solidification structures.
\begin{figure}[t]
    \centering
    \includegraphics[scale=0.37]{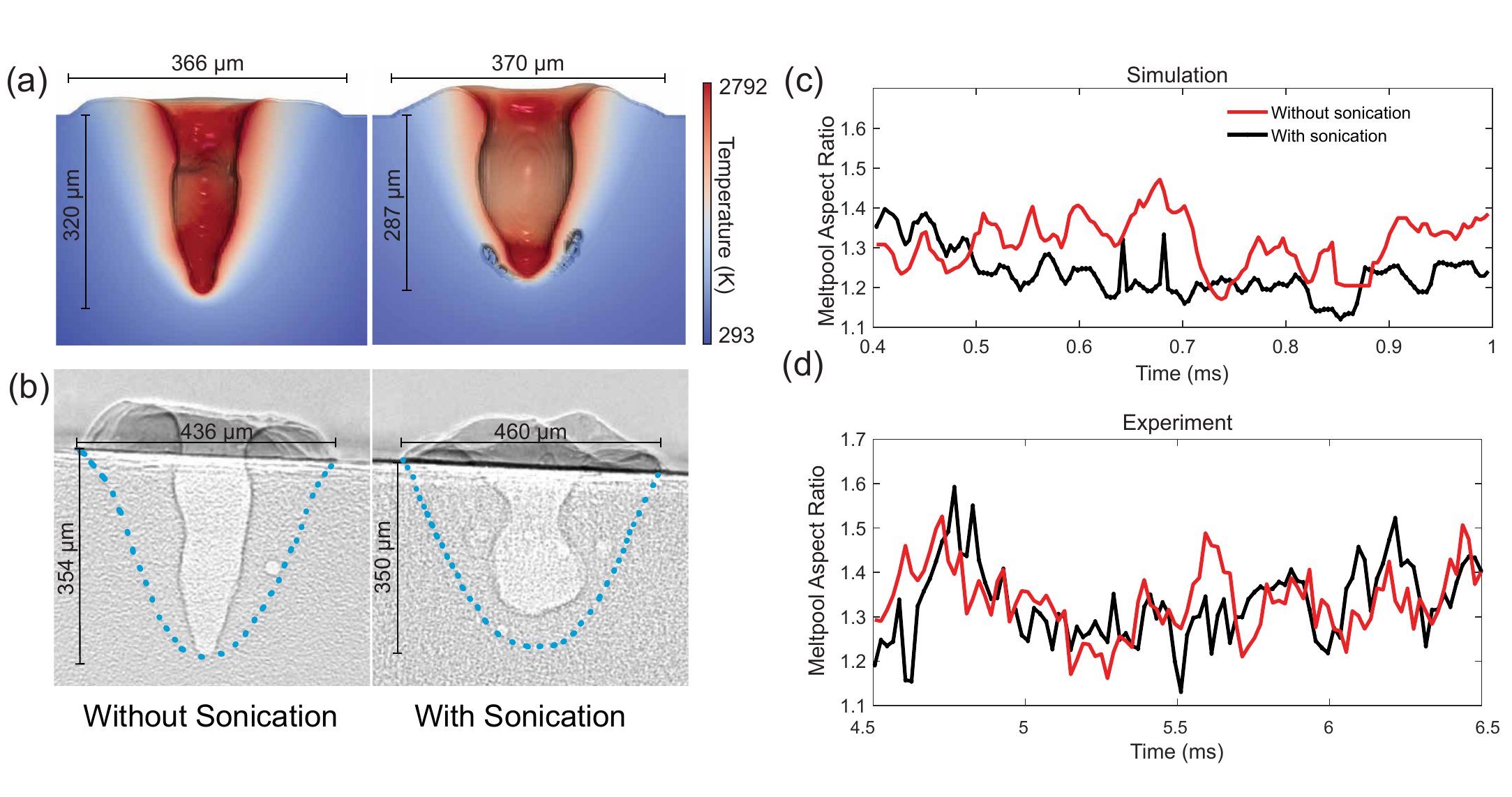}
    \caption{\textbf{CFD melt pool simulation comparison with X-ray results.} (a) Melt pool simulation without and with sonication, (b) comparable experimental results without and with sonication, (c) Aspect ratios (depth/width) observed in the simulations, and (d) corresponding experimental aspect ratios.}
    \label{fig:MeltPoolEXP&SIM}
\end{figure}
Figures \ref{fig:MeltPoolEXP&SIM} (a) and (b) compare CFD simulated melt pools to melt pool geometries directly captured in X-ray imaging for the cases of without and with sonication. Deep and narrow keyholes observed with high-speed X-ray imaging in melt pools without sonication were replicated in the simulations. Similarly, an increased keyhole tip radius observed with X-ray imaging in melt pools when sonication was included in the Flow-3D\textcircled{\scalebox{0.7}{R}} simulation. Supplementary Movies 3 and 4 show simulated keyhole dynamics for the two cases. Furthermore, Supplementary Movies 5 and 6 show results of simulated melt pool dynamics. Similar to the melt pool dynamics undergoing sonication captured by X-ray imaging (i.e. Supplementary Movie 2), the simulated melt pools (i.e. Supplementary Movie 5) showed acoustic cavitation-driven bubble nucleation and implosion caused by pressure variation in the melt pool. Furthermore, the simulated solidification structure with ultrasonic treatment shows frozen cavitation-induced pores like those observed in X-ray imaging and X-ray CT. To further validate the simulations, the measured melt pool aspect ratios (width/depth) from x-ray images were compared with the simulated melt pool aspect ratios. Figures \ref{fig:MeltPoolEXP&SIM} (c) and (d) show melt pool aspect ratios, which were found to be closely consistent between simulations and experiments. The close agreement in aspect ratios speaks to the simulations accurately representing the laser energy transfer into the pool

\medskip

\begin{figure}[h]
    \centering
    \includegraphics[scale=0.35]{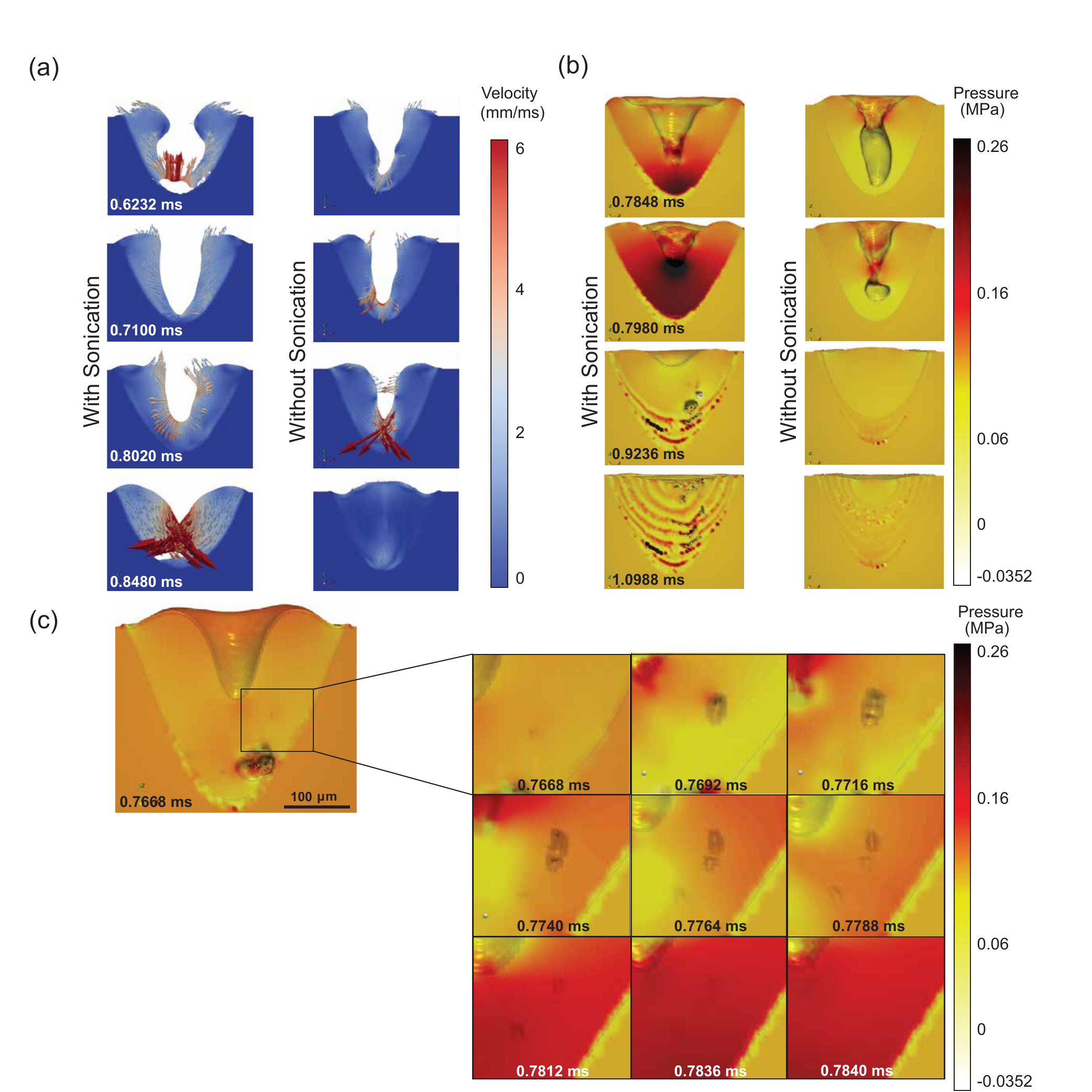}
    \caption{\textbf{CFD melt pool simulation results with and without sonication.} (a) Simulation frames showing velocity vectors of points in the liquid, (b) pressure distributions, (c) pressure field at the nucleation of a cavitation bubble and after the collapse.}
    \label{fig:Vol_Pressure_Distribution}
\end{figure}

\noindent Melt pool flow dynamics are primarily driven by surface tension, Marangoni convection, and recoil pressure. The application of ultrasound introduces acoustic streaming as an additional driving force. Simulations allowed us to quantify acoustic streaming by comparing velocity vectors at points in the fluid with and without applied ultrasonic treatment. Figure \ref{fig:Vol_Pressure_Distribution} (a) shows melt pool speed contours and velocity vectors with and without sonication, respectively. Supplementary Movies 7 and 8 show additional melt pool dynamics. The higher melt pool velocities in melt pools with sonication confirm that acoustic streaming is a major factor in fluid flow.
\noindent Figure \ref{fig:Vol_Pressure_Distribution} (b) shows the pressure distribution. Large pressure fluctuations are observed in the presence of sonication. The frames shown in Figures \ref{fig:Vol_Pressure_Distribution} (b) were taken from the simulation results during solidification and when the laser was switched off. This was done to decouple the sonication from thermal energy input. Supplementary Movies 9 and 10 show animations of pressure distribution in solidifying melt pools with and without sonication, respectively. It can be seen from Figure \ref{fig:Vol_Pressure_Distribution} (b) that the pressure variation in the melt pool with sonication promoted bubble nucleation. In addition, the influence of ultrasonic vibration can be observed in Figure \ref{fig:Vol_Pressure_Distribution} (b) with sonication, as ripples of high and low-pressure regions captured by the solidification. Without sonication, no significant pressure variation was observed during solidification. Acoustic cavitation bubble nucleation occurs when the localized pressure within a liquid drops below the vapor pressure of that liquid. Therefore, in Al6061 laser-generated melt pools, it can be seen that if the localized pressure within the melt pool drops below the vapor pressure of molten Al6061, nucleation of bubbles will occur. To investigate the influence of pressure variation on bubble nucleation during melting, the image sequence in Figure \ref{fig:Vol_Pressure_Distribution} (c) shows the pressure contours at a bubble nucleation site within the melt pool. A decrease in melt pool pressure was observed to result in bubble nucleation, while an increase in pressure promoted bubble implosion.

\medskip

\noindent Microstructure development is directly linked to solidification rates and thermal gradients. To investigate the influence of ultrasonic treatment on solidification conditions, we collected time history temperature gradients and cooling rates at a point within the melt pools with and without sonication.
\begin{figure}[h]
    \centering
    \includegraphics[scale=0.2]{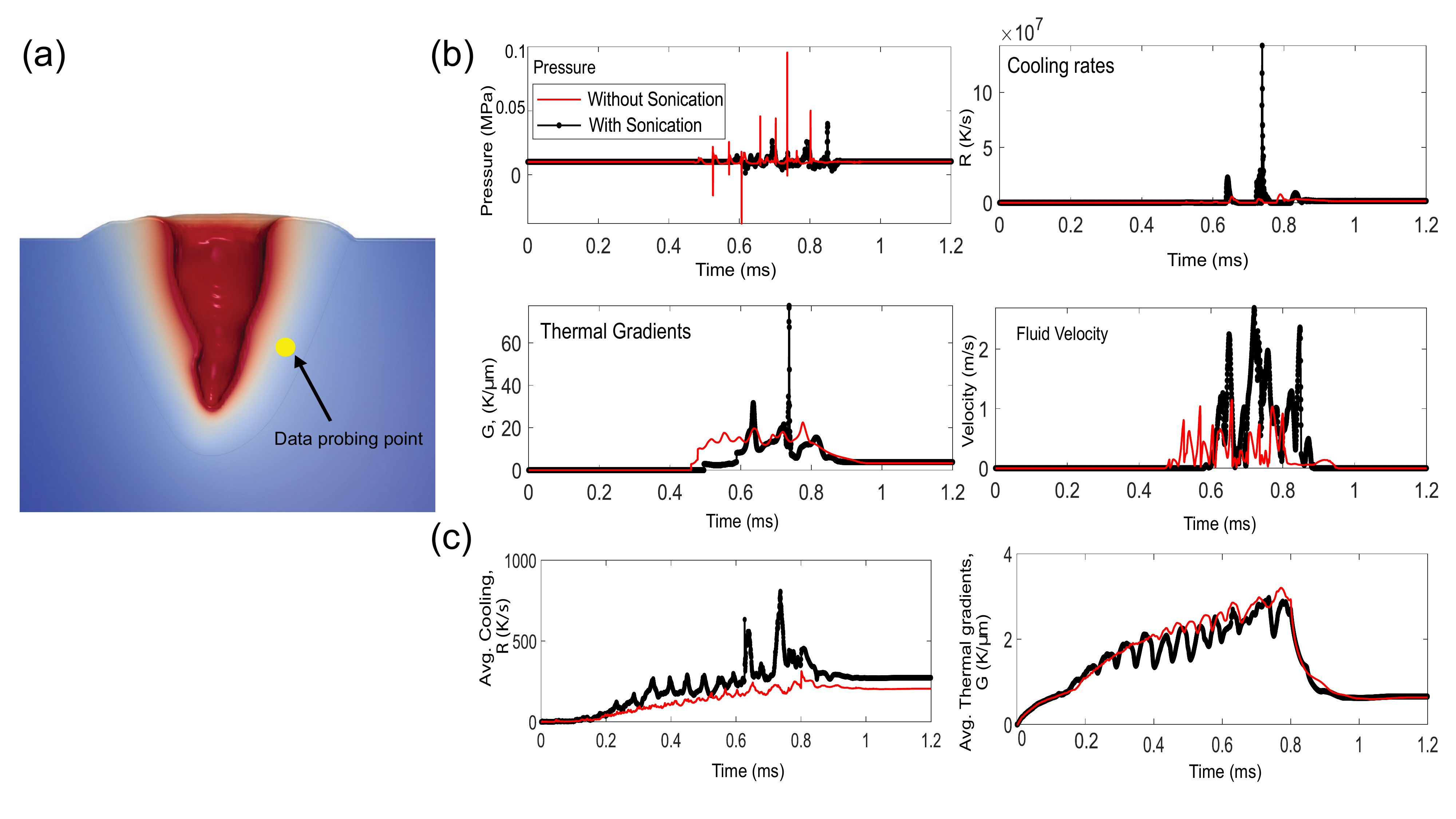}
    \caption{\textbf{Melt pool thermal history from CFD simulation.} (a) Point data probing location. (b) The time history of fluid pressure, cooling rates, thermal gradients, and fluid velocities at the data probing point with and without sonication. (c) Melt pool thermal gradients and cooling rates at each time frame during the entire simulation with (red line plain line) and without sonication (black line with circular markers).}
    \label{fig: Cooling rates and thermal gradients}
\end{figure}
\noindent Figure \ref{fig: Cooling rates and thermal gradients} (a) shows the point data probing location at which the time history of parameters that can be related to microstructural development was collected. Figure \ref{fig: Cooling rates and thermal gradients} (b) shows the time history of pressure, cooling rate, thermal gradient, and velocity at the data probing point, with and without sonication. It can be observed that high pressure was observed in melt pools without compared to those with sonication. Conversely, higher cooling rates were observed in melt pools with sonication. Similarly, higher thermal gradients and fluid velocities were observed in melt pools with compared to those without sonication. Figure  \ref{fig: Cooling rates and thermal gradients} (c) shows the overall cooling rates and thermal gradients at each simulation time frame over the entire simulation. It can be observed that the overall thermal gradient did not respond to ultrasonic treatment. However, the overall melt pool cooling rate increased with the applied ultrasonic treatment.

\medskip

\begin{figure} [t]
    \centering
    \includegraphics[width=1\textwidth]{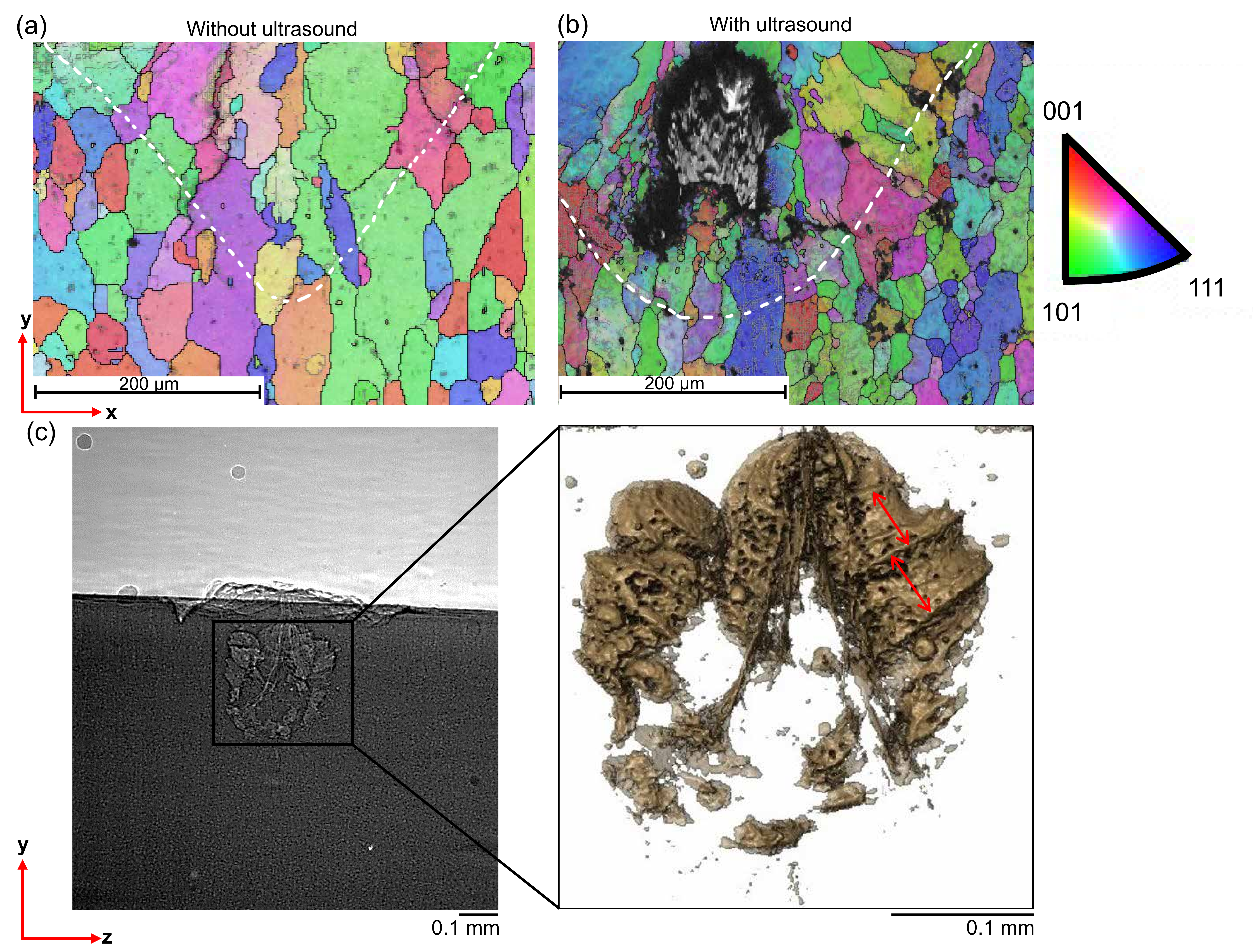}
    \caption{\textbf{Melt pool EBSD and X-ray computed tomography analysis results.} EBSD grain map showing the solidification microstructure (a) without and (b) with sonication. (c) High-speed X-ray frame showing the final solidification structure and corresponding X-ray computed tomography visualization showing the porosity features and indications of the sonication-driven vibrations (seen by the red arrows).}
    \label{fig: X-ray CT scan}
\end{figure}

\noindent \textbf{Acoustic cavitation characterization and influence on microstructural development.} The primary aim of this article is to unveil the physics associated with ultrasonically driving the melt pool. A secondary aim and a topic of future work is to unveil conditions that lead to refined or tailored microstructures toward improved quality and performance of AM parts. Nevertheless, the solidification microstructures formed in melt pools with and without sonication were characterized using electron backscatter diffraction (EBSD).  Figures \ref{fig: X-ray CT scan} (a) and (b) show the microstructures and crystallographic orientations of the grains in melt pools without and with sonication, respectively. Since EBSD is destructive, it is noted that the non-sonicated case is a different sample having a single point melt with the same laser power and duration as the sonicated melt pool case. For both samples, the melt pool boundary was traced using standard SEM images, in which the melt region was clear (see supplementary document) and then superimposed on the EBSD grain map. Epitaxial grain growth and cracking along grain boundaries were evident in both cases. A qualitative reduction in grain size is observed in the sonication case but is difficult to ascertain because of the large pore structure as seen by the dark features in Figure \ref{fig: X-ray CT scan} (b)).

\medskip

\noindent Moreover, X-ray computed tomography analysis was performed on the final solidification structure (prior to EBSD) to characterize the influence of cavitation and acoustic streaming in sonicated laser-generated melt pools. An X-ray frame from the high-speed imaging showing the final solidification structure and a 3D isosurface of cavitation-induced porosity in the melt pool is shown in Figure \ref{fig: X-ray CT scan} (c). The X-ray computed tomography reveals evidence of frozen cavitation bubbles and ultrasonic vibration-induced ripples in the melt pool (i.e. labeled by arrows in the X-ray computed tomography scan image). The ultrasonic wavelength in Al6061 at a frequency of $20.2$ kHz was calculated to be $0.32$ m, which is orders of magnitude higher than the melt pool depth and width. Thus, the micron scale ripples observed resulted from the sinusoidal variation in pressure from the ultrasonic vibration, which we have also observed in CFD simulations. This discovery calls attention to the influence of vibration amplitudes on cavitation in laser-based AM with ultrasonic treatment, which has not been previously explored. Figure \ref{fig: X-ray CT scan} (c) also shows a higher concentration of pores near the sample surface relative to the bottom of the melt pool. Thus, it is further corroborated that ultrasonic treatment causes bubble migration toward the melt pool surface.

\medskip

\noindent Cavitation in ultrasonic molten metal processing has been explored by several researchers \cite{Eskin98, Eskin2018chp5, Eskin19, Tzanakis20}, who conducted casting experiments on light metallic alloys. High-temperature cavitometry \cite{XU2023} and high-speed imaging \cite{TZANAKIS2015} were used to establish a cavitation threshold in terms of acoustic intensity \cite{Atchley89}. The first-order linear approximation of ultrasonic intensity, $I$, in an acoustic field is \cite{Eskin98}
\begin{equation}
    I = \frac{1}{2} \rho c (2 \pi A f)^2, 
    \label{Eq:1}
\end{equation}
where $\rho$ is the fluid density, $c$ is the speed of sound in the fluid, $A$ is the wave amplitude and $f$ is the ultrasonic frequency. An acoustic intensity cavitation threshold of $100$ W/cm$^2$ was established for light metal alloys through casting experiments with ultrasonic melt processing \cite{Eskin98}. In the experiments described in the literature \cite{WangB18, Tzanakis20}, an ultrasonic transducer horn was immersed in molten metal to introduce a propagating wave directly into the solidifying metal. Hence, the cavitation threshold could be established for sizable molten metal pools, and solidification rates would be significantly lower than those in AM processes. Nevertheless, the  $100$ W/cm$^2$ cavitation threshold has been proposed for laser-based AM printing of light metallic alloys on sonicated substrates \cite{Todaro20, Todaro2021, Yang22, Feilong2023, Ji2023, Yuan2021, Xu2022, Ning2018}. However, laser AM fundamentally differs from casting because of the submillimeter-size melt pools that exist for milliseconds owing to the associated rapid solidification rates. In casting, metal melting and solidification are separate processes, whereas melting, molten metal agitation and solidification occur simultaneously in laser AM with sonication to generate acoustic cavitation. In addition, ultrasonic melt processing in casting involves wave propagation in a solidifying molten metal, while in AM, it involves local vibration of the molten metal. Such factors indicate different physical environments for cavitation in casting and AM. Therefore, validation of acoustic cavitation thresholds in laser-generated melt pools is needed, underpinning the importance of our technique.

\medskip

\noindent Using a wave speed of $4718$ m/s, density of $2586$ kg/m$^{3}$, wave amplitude of $8$ $\upmu$m, and frequency of $20.2$ kHz in Equation (1) resulted in an acoustic intensity of $628.9$ W/cm$^{2}$. Our calculated acoustic intensity is above the established $100$ W/cm$^2$ cavitation threshold. However, cavitation was observed in the CFD simulations at an average acoustic intensity of $10$ W/cm$^2$, which is much lower than the established cavitation intensity threshold and the calculated intensity from Equation (1). Therefore, the established cavitation threshold from casting light metals with sonication overestimates the acoustic intensity required to induce cavitation in laser-generated melt pools on vibrating substrates. In the future, we will explore the influence of acoustic intensity on cavitation, porosity, and microstructure refinement.

\medskip

\section*{Discussion}
\noindent The application of ultrasound in solidifying melt pools in laser-based AM has been shown to promote columnar to equiaxed grain transition \cite{Wang2023} resulting in improved and homogenized mechanical properties and random crystallographic orientations \cite{Yuan2021}. By adopting observed microstructural refinement mechanisms in casting with ultrasonic treatment, acoustic cavitation and streaming \cite{Eskin19} have been hypothesized as the primary driving mechanisms of microstructural refinement in laser-based AM. Unambiguous evidence of cavitation in sub-millimeter scale and opaque laser-generated melt pools has been elusive until now. Here, the real-time influence of ultrasonic vibration on melt pool, keyhole, and bubble dynamics and the solidification of laser-generated melt pools was revealed. We also elucidated the impact of ultrasonic vibration at $20.2$ kHz on melt pool and keyhole morphologies. Furthermore, we explained the potential influence of ultrasonic vibration on laser energy absorptivity and its benefits in AM. EBSD and XCT techniques were used to analyze the microstructures and solidification structures with and without applied sonication. The influence of ultrasonic vibration on melt pool flow velocity, pressure distribution, and solidification conditions with and without sonication was investigated using Flow-3D\textcircled{\scalebox{0.7}{R}} multiphysics CFD simulation software.

\medskip

\noindent Melt pool and keyhole dynamics in laser-based AM processes influence porosity formation mechanisms \cite{Hojjatzadeh2019} and dictate the resulting solidification microstructures \cite{Tan2015} and mechanical properties \cite{Chen2022} of AM components. Marangoni flow, recoil pressure, and surface tension are some of the major driving forces governing melt pool and keyhole dynamics \cite{Leung2018}. Generating melt pools on a substrate vibrating at ultrasonic frequencies introduces an additional force that drives melt pool flow in the wave propagation direction (i.e. acoustic streaming) \cite{Yang22}, which competes with existing forces in the melt pool. We used high-speed synchrotron X-ray imaging and Flow-3D\textcircled{\scalebox{0.7}{R}} simulations to show that acoustic streaming dominates the melt pool and keyhole dynamics in the laser-generated melt pool with sonication. Moreover, physical evidence of real-time acoustic cavitation in submillimeter-sized laser-generated melt pools was revealed in situ using high-speed X-ray imaging. Ultrasonic vibration was observed to increase bubble density in the melt pool and promote bubble migration toward the melt pool surface. X-ray computed tomography scan of the final solidification structure further demonstrated that ultrasonic vibration drives pores toward the melt pool surface and that vibration amplitude influences molten metal flow rather than ultrasonic wavelength.

\medskip

\noindent Keyhole morphology analysis from high-speed X-ray images revealed a wide and shallow keyhole with applied sonication. A deep and narrow keyhole was observed in the case without sonication. Deep and narrow keyhole geometries are susceptible to keyhole tip collapse porosity \cite{Zhao2020}; therefore, by changing the keyhole morphology, ultrasonic treatment could potentially eliminate one of the major porosity formation mechanisms in laser AM. It is important, however, to note that sonication-induced cavitation resulted in porosity, as revealed by post-process EBSD and X-ray computed tomography scan results Therefore, these observations spark interest in further investigations on ultrasonic wave parameter optimization to leverage cavitation for porosity reduction and location-specific microstructural refinement. Furthermore, cavitation-induced porosity in AM ultrasonic melt processing could be used to manufacture porous structures for biomedical applications. Frequency modulation and the use of multiple ultrasound sources could potentially provide a certain degree of control over cavitation in laser-generated metal pools.

\medskip

\noindent The application of ultrasonic vibration in laser-based AM was considered to increase the laser beam reflection from the liquid/gas interface in the melt pool because of increased keyhole depth caused by the raised keyhole rim. Increased laser beam reflection can potentially improve laser energy absorptivity \cite{Simonelli2015}, resulting in larger melt volumes. On the other hand, applying ultrasonic treatment through out-of-plane vibration increased hot spattering due to the molten metal droplets pinching off the melt pool at peak positive and negative vibration amplitudes. Further optimizing vibration frequency, amplitudes, and direction can help mitigate hot spattering.

\medskip

\noindent To investigate the influence of ultrasonic treatment on solidification and microstructural development, we utilized Flow-3D\textcircled{\scalebox{0.7}{R}} multiphysics simulations validated with real-time high-speed synchrotron X-ray images of melt pool dynamics. Flow-3D\textcircled{\scalebox{0.7}{R}} simulation results showed pressure variation-driven acoustic cavitation in melt pools with applied ultrasonic treatment. The pressure variation in melt pools with and without applied ultrasound was analyzed during the solidification phase (i.e. after the laser was switched off) using color maps. Ultrasonic treatment was also observed to promote high melt pool velocities, cooling rates, and thermal gradients. Higher thermal gradients and melt pool velocities create stronger cooling effects and promote heterogeneous nucleation and grain refinement.

\medskip

\noindent \noindent In summary, we provided evidence of acoustic cavitation in laser-generated molten metal pools on sonicated substrates using both high-speed X-ray imaging and CFD simulations. We further showed that ultrasonic treatment influenced melt pool and keyhole dynamics and could potentially eliminate some major keyhole porosity driving mechanisms. We also demonstrated through simulations that ultrasonic treatment creates favorable conditions for heterogeneous nucleation and grain refinement. These results facilitate further investigation into the influence of ultrasonic treatment on microstructural refinement and mechanical property improvement in laser-based AM processes.

\medskip

\section*{Methods}
\subsection*{Materials and sample preparation}
Al6061 alloy was chosen as the material of interest because of its widespread usage in lightweight material industries such as automotive, aerospace, and many others. Unfortunately, Al6061 is extremely challenging to use in welding or AM because of thermal cracking. Thus, this research has a broader goal of investigating techniques to improve the printability of such alloys. Moreover, manufacturing methods, processes, and conditions highly influence Al6061 grain sizes and mechanical properties, as demonstrated by Eskin \cite{Eskin98} in the ultrasonic treatment of light metallic alloys. Secondly, applications of Al6061 as an additive manufacturing material have been limited because of residual stress build-up \cite{Mehta2021}. Lastly, Al6061 has liquidus and solidus temperatures of 660 C and 595 C, respectively, enabling sizable mushy zones necessary for effective and efficient ultrasonic treatment. Al6061 samples with a length of $20$ mm, a height of $12$ mm, and a thickness of $1.5$ mm were used in our experiments. A thickness of $1.5$ mm allowed adequate X-ray absorption contrast between the solid, liquid, and gaseous phases during laser melting, making it easy to identify melt pool features (i.e. vapor/gas depression, bubbles, solid-liquid interfaces.).

\medskip

\subsection*{Ultrasonic wave generation system}
Al6061 specimens were adhered to an ultrasonic transducer horn using an adhesive, as illustrated in Supplementary Figure 1. A $20.2$ kHz high-power ultrasonic transducer by Hangzhou Altrasonic Technology Co., Ltd., with a maximum power of $2000$ W, was used in this study. The ultrasonic system consisted of a horn, piezoelectric elements, and an ultrasonic generator. The ultrasonic generator converts the power source into high-frequency and high-voltage alternating current and transmits it to the piezoelectric elements, which convert the input electrical energy into mechanical energy (i.e. ultrasonic waves). In our experiments, the transducer generated a continuous longitudinal wave and was operated at the horn's resonant frequency of $20.2$ kHz, with a power of $600$ W and vibration amplitude of $8$ $\upmu$m. The transducer power and short time intervals of ultrasonic wave application were chosen to prevent the transducer horn overheating, which may influence melt pool solidification rates and thermal gradients. A custom-designed relay apparatus operated from outside the experimental hutch controlled the transducer on/off switching and the duration of ultrasonic vibration.

\medskip

\subsection*{X-ray imaging and laser melting system}
Experiments were conducted using the high-energy ultrafast synchrotron X-ray imaging system available at the Advanced Photon Source, Argonne National Laboratory, USA. The 32-ID-B beamline at the Advanced Photon Source offers a state-of-the-art high-speed X-ray imaging technique. The intense undulator white beam allows ultrafast image acquisition rates of $50$ kHz with a spatial resolution of $2$ $\upmu$m/pixel in a field of view of $1.8\times1$ mm. In addition, a continuous-wave ytterbium fiber laser (IPG YLR-$500$-AC, IPG Photonics, Oxford, USA, wavelength of $1070$ nm, maximum output power of $560$ W) and a galvano scanner (IntelliSCANde $30$, SCANLAB GmbH., Germany) \cite{Hojjatzadeh2019} were integrated to perform stationary laser melting on bare Al6061 samples. A laser power of $350$ W was used in the experiments. Experiments were conducted in the following sequence: the X-ray shutter and camera were first opened to initiate image acquisition. Secondly, the ultrasonic transducer was switched on, and lastly, the laser was turned on. The experimental setup and sequence allowed the sample melting, vapor depression development, and melt pool solidification occurring in an acoustic field to be captured via X-ray imaging. The laser was switched on for $3.34$ ms for both cases with and without ultrasonic treatment.

\medskip

\subsection*{ EBSD and X-ray computed tomography analysis}
Electron backscattered diffraction patterns (EBSPs) were obtained in the Oxford scanning electron microscope (SEM) instrument by focusing an electron beam on the Al6061 sample. The final polishing of the Al6061 sample was conducted using the Final A polishing pad with $0.04$-micron colloidal silica suspension for 12 hours. The sample was tilted to approximately 70 degrees with respect to the horizontal, and the diffraction patterns were imaged on a phosphor screen. The images were captured using a low-light CMOS camera. A $1.5$-micron step size was used for both samples with and without ultrasonic treatment. The X-ray computed tomography scan was conducted with a Zeiss Xradia Versa $620$ CT system using a source accelerating voltage of $80$ kV. Images were acquired over $2$ hours at a voxel size of $1.5$ $\upmu$m and reconstructed using Zeiss proprietary software. The dicom image files were then processed using MATLAB to reveal the influence of ultrasonic vibration on the final solidification structure of the melt pools. A 3D view of sonication-induced pores showing the influence of ultrasonic vibration amplitude on the melt pool solidification was captured using the 3D volume viewer tool in MATLAB. 

\medskip

\subsection*{Image processing}
MATLAB image processing toolkit and ImageJ were utilized in the X-ray image analysis. MATLAB codes were developed to normalize a sequence of X-ray images with their average pixel values. To create an X-ray image sequence with a uniform gray value, images within a $5\%$ range of gray values were grouped together. A normalization operation was applied to each distinct group, which allowed enhanced visualization of melt pool features, keyhole dynamics, and bubble motion. Measurements of the melt pool and keyhole depth changes and bubble motion characterization were conducted using ImageJ. Maximum depths and widths on each X-ray frame measured in ImageJ were used to characterize melt pool and keyhole dynamics. The peak-to-peak vibration amplitude on the Al6061 sample surface was also measured as $16$ $\upmu$m using ImageJ. 

\medskip

\subsection*{Multiphysics computational fluid dynamic simulations}
A $1$ mm$^2$ domain with a 4-micron mesh size was used in the CFD simulations on the Flow-3D\textcircled{\scalebox{0.7}{R}} platform. The simulation finish time was set at $1.3$ ms, and the laser on time was set at $0.8$ ms. Similar to our experiments, the laser power used in the simulations was $350$ W, with a laser spot size of $80$-microns. Ultrasonic vibration was introduced by defining a non-inertial reference frame with harmonic oscillations on the melt volume (i.e. substrate). The oscillation frequency was set at $20.2$ kHz and an amplitude of $8$-microns. The execution time for each simulation with and without ultrasonic treatment was one day and $16$ hours, with each model generating a $2.5$ TB output data file (More details on the simulation setup, boundary conditions, and governing equations are provided in the supplementary document).

\medskip

\section*{Acknowledgments}
This research used the Advanced Photon Source resources, a U.S. Department of Energy (DOE) Office of Science user facility operated for the DOE Office of Science by Argonne National Laboratory under Contract No. DE-AC02-06CH11357. In addition, this work was partially supported by the US DOE LLNL-LDRD 23-ERD-009 under the auspices of the U.S. Department of Energy by Lawrence Livermore National Laboratory under Contract DE-AC52-07NA27344. Lastly, we acknowledge Dr. Zhongshu Ren and Dr. Jiayun Shao from Northwestern University for their valuable contributions to the computational fluid dynamic simulations we conducted using the Flow-3D \textcircled{\scalebox{0.7}{R}} additive manufacturing software package.

\section*{Author contributions}
Lovejoy Mutswatiwa conducted electron backscattered difference analysis and wrote the manuscript with the help of Andrea Paola Argüelles, Judith Anne Todd, and Christopher Micheal Kube. Lauren Katch, Nathan John Kizer, Tao Sun, Samuel James Clark, Kamel Fezzaa, Jordan Lum, and David Matthew Stobbe facilitated securing beam time at the 32-ID-B beamline at the Advanced Photon Source, Argonne National Laboratory, USA. These authors also helped set up the high-speed synchrotron X-ray imaging experiments and operate the 32-ID-B beamline. Griffin Jones and Kenneth Charles Meinert Jr. helped with X-ray computed tomography analysis.

\section*{Data availability}
Data supporting the findings presented in this study are available from the corresponding author upon request.

\section*{Competing interests}
The authors declare no competing interests.

\section*{Additional Information}
\textbf{Supplementary information} is provided to elucidate and further support the reported findings in this study.


\end{document}